\documentclass[conference]{IEEEtran}
\IEEEoverridecommandlockouts

\usepackage{amsmath,amssymb}
\usepackage{graphicx}
\usepackage{cite}
\usepackage{booktabs}
\usepackage{hyperref}
\usepackage{fancyhdr} 
\begin{document}

\title{RTR: A Transformer-Based Lossless Crossover with Perfect Phase Alignment}

\author{
\IEEEauthorblockN{Xiangying Li\textsuperscript{1,*}\textsuperscript{,} \quad Jiankuan Li\textsuperscript{2,}\textsuperscript{\textdagger}\textsuperscript{,} \quad Yong Tang\textsuperscript{3,}\textsuperscript{\textdagger}}
\IEEEauthorblockA{\textsuperscript{1}School of Computer, South China Normal University, China \\
\textsuperscript{2}School of Health, Leeds Beckett University, UK \\
\textsuperscript{3}School of Computer, South China Normal University, China \\
Email: lixiangying@pku.edu.cn }
}

\maketitle
\thispagestyle{plain} 
\footnotetext{
  \copyright 2025IEEE.  Personal use of this material is permitted. Permission from IEEE must be obtained for all other uses, in any current or future media, including reprinting/republishing this material for advertising or promotional purposes, creating new collective works, for resale or redistribution to servers or lists, or reuse of any copyrighted component of this work in other works.}
\IEEEpubidadjcol

\begin{abstract}
This paper proposes a transformer-based lossless crossover method, termed \emph{Resonant Transformer Router} (RTR), which achieves frequency separation while ensuring perfect phase alignment between low-frequency (LF) and high-frequency (HF) channels at the crossover frequency. The core property of RTR is that its frequency responses satisfy a linear complementary relation
\[
H_{\mathrm{LF}}(j\omega)+H_{\mathrm{HF}}(j\omega)=1,
\]
so that the original signal can be perfectly reconstructed by linear summation of the two channels. Theoretical derivation and circuit simulations demonstrate that RTR provides superior energy efficiency, phase consistency, and robustness against component tolerances. Compared with conventional LC crossovers and digital FIR/IIR filters, RTR offers a low-loss, low-latency hardware-assisted filtering solution suitable for high-fidelity audio and communication front-ends.

The core theory behind this paper's work, lossless crossover, is based on a Chinese patent [CN116318117A] developed from the previous research of one of the authors, Jiankuan Li. We provide a comprehensive experimental validation of this theory and propose a new extension.
\end{abstract}


\section{Introduction}
Frequency splitting and signal reconstruction are fundamental operations in audio processing, communications, and array signal processing. Conventional passive LC crossovers, while simple to implement, suffer from energy loss, passband ripple, and phase mismatch around the crossover frequency. As documented in prior research, such as \cite{Cao2022}, the performance of these LC-based designs is highly sensitive to component tolerances, a critical issue in real-world applications. 

While digital filters provide a high degree of design flexibility, they often introduce significant computational and latency overhead. In high-fidelity audio applications, as highlighted by \cite{Huang2024}, the challenges of improving clarity and fidelity in complex environments through adaptive filtering algorithms remain a key area of research, particularly concerning computational complexity.

To overcome these issues, we propose a novel transformer-based method for the automatic routing of low- and high-pass signals, which we call the Resonant Transformer Router (RTR). The core idea is to exploit transformer coupling to form complementary LF and HF branches whose magnitude responses sum linearly to unity while maintaining strict 0¡ã phase alignment. This unique approach, which is inherently more robust to component variations than traditional methods, ensures the original signal can be reconstructed without distortion by direct analog summation of the two outputs.

Our main contributions are as follows.
\begin{itemize}
  \item Propose and analytically derive a transformer-based crossover and reconstruction model;
  \item Prove energy conservation and linear-phase alignment under ideal assumptions and analyze non-ideal error terms;
  \item Validate performance via circuit simulation and compare with LC and digital filters;
  \item Discuss applications in audio, RF front-ends, array processing and network-transport security scenarios.
\end{itemize}

\section{RTR Model and Theoretical Analysis}
\subsection{Basic structure and linear complementarity}
\begin{figure}[h]
  \centering
  \includegraphics[width=0.9\linewidth]{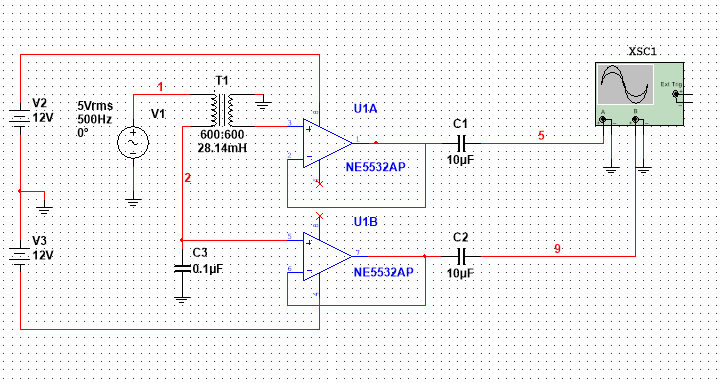}
  \caption{RTR test circuit}
  \label{fig:circuit}
\end{figure}

As can be seen in Fig.~\ref{fig:circuit}, the topology of the RTR consists of a transformer T1 and a capacitor C3. The capacitor is connected in series with the primary coil and the junction outputs a low-frequency signal (LPF). The secondary side of the transformer outputs a high-frequency signal (HPF). 

Let the input voltage be \(V_{\mathrm{in}}(j\omega)\). The LF and HF outputs taken from the appropriate nodes are \(V_{\mathrm{LF}}(j\omega)\) and \(V_{\mathrm{HF}}(j\omega)\). Define transfer functions
\[
H_{\mathrm{LF}}(j\omega)=\frac{V_{\mathrm{LF}}(j\omega)}{V_{\mathrm{in}}(j\omega)},\quad
H_{\mathrm{HF}}(j\omega)=\frac{V_{\mathrm{HF}}(j\omega)}{V_{\mathrm{in}}(j\omega)}.
\]
Under ideal transformer assumptions (perfect coupling, negligible winding resistance, and parasitics) and with high-impedance buffering at outputs, the secondary-derived HF voltage satisfies
\[
V_{\mathrm{HF}}(j\omega)=V_{\mathrm{in}}(j\omega)-V_{\mathrm{LF}}(j\omega).
\]
Therefore,
\begin{equation}
H_{\mathrm{HF}}(j\omega)=1-H_{\mathrm{LF}}(j\omega),
\label{eq:complement}
\end{equation}
and equivalently
\begin{equation}
H_{\mathrm{LF}}(j\omega)+H_{\mathrm{HF}}(j\omega)=1.
\label{eq:sum1}
\end{equation}
Equation~\eqref{eq:sum1} ensures that for every frequency the two channel voltages sum to the input:
\[
V_{\mathrm{LF}}(j\omega)+V_{\mathrm{HF}}(j\omega)=V_{\mathrm{in}}(j\omega),
\]
so in the time domain the original signal is recovered: \(x(t)=x_{\mathrm{LF}}(t)+x_{\mathrm{HF}}(t)\).

\subsection{Non-ideal effects}
When accounting for winding resistance \(R_w\), leakage inductance \(L_\sigma\), finite coupling coefficient \(k\) and parasitic capacitances, the ideal relation in \eqref{eq:sum1} acquires a small error term:
\begin{equation}
H_{\mathrm{LF}}(j\omega)+H_{\mathrm{HF}}(j\omega)=1+\varepsilon(j\omega),
\label{eq:nonideal}
\end{equation}
where \(\varepsilon(j\omega)\) is frequency dependent and ideally small. Sources of \(\varepsilon\) include imperfect coupling (leakage), resistive losses and loading effects. With careful transformer design and output buffering, \(|\varepsilon(j\omega)|\) can be made sufficiently small for engineering needs (e.g., $<-40\,$dB across the passband).

\subsection{Energy conservation and resonance}
As a passive, reactive network, RTR obeys energy conservation:
\[
E_{\mathrm{in}}(t)=E_{\mathrm{LF}}(t)+E_{\mathrm{HF}}(t).
\]
At the crossover frequency, an amplitude peak may appear in both the LF and HF branches as a result of LC series resonance formed by the primary inductance and capacitance. At resonance the net impedance is minimized, maximizing energy transfer to the outputs. This peak represents efficient transfer rather than energy dissipation, and thus does not violate the law of conservation. Crucially, because the low-frequency (LF) and high-frequency (HF) signals remain in phase at all times, once a load or its corresponding notch filter is connected, their linear superposition yields a nearly flat overall response.

\section{Simulation and Experimental Results}
In order to comprehensively evaluate the lossless performance of the RTR, we compared its output signals with the original input signals through simulation. A key challenge in objective audio quality assessment is the time delay and synchronization between the reference and test signals, as highlighted by prior research \cite{Chang2010}. Owing to the RTR's strict 0¡ã phase alignment property, its reconstructed signal maintains perfect phase coherence with the original signal. This allows us to perform direct waveform comparison without the need for complex digital synchronization algorithms, thereby significantly simplifying the performance evaluation process.

\subsection{Simulation setup}
\begin{figure}[h]
  \centering
  \includegraphics[width=0.9\linewidth]{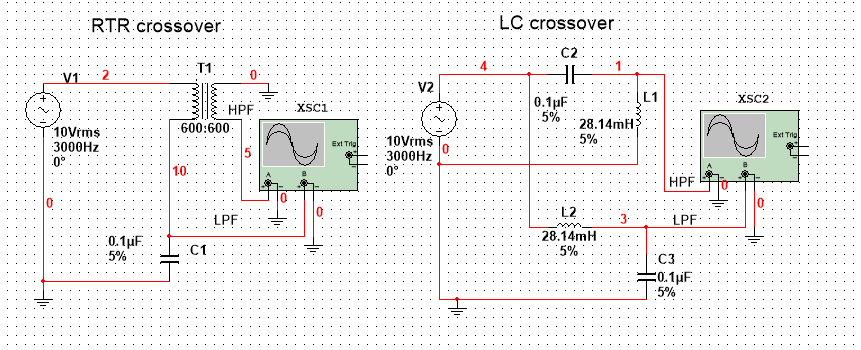}
  \caption{RTR and LC crossover}
  \label{fig:rtr_lc}
\end{figure}

The RTR circuit was implemented in Multisim.Fig.~\ref{fig:rtr_lc} A second-order LC crossover and a representative FIR implementation were built for comparison. Input signals included multi-tone sinusoids and broadband audio sweeps. Measured metrics: amplitude response, phase response, insertion loss, power flow and tolerance sensitivity.\\

\subsection{Amplitude and phase characteristics}
Simulation results show that the LF and HF amplitude responses meet the linear complementarity relation \((H_{\mathrm{LF}}+H_{\mathrm{HF}}=1)\) within numerical precision. Phase traces for both branches remain aligned at 0¡ã across the operating band, while the LC crossover exhibits notable phase deviations (10¡ã--30¡ã) near the crossover.

\begin{figure}[h]
  \centering
  \includegraphics[width=0.7\linewidth]{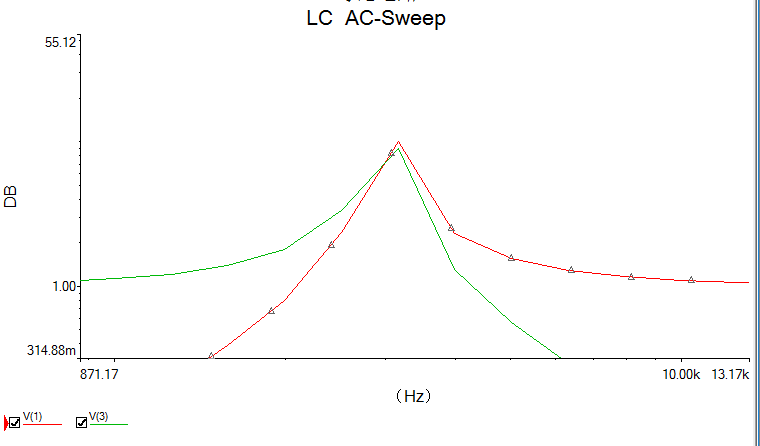}
  \caption{LC crossover AC-sweep}
  \label{fig:lc_amplitude}
\end{figure}

As can be seen from the Fig.~\ref{fig:lc_amplitude}, under the premise of a 5\% component tolerance, the resonant frequencies of the high-pass filter (HPF) and low-pass filter (LPF) in the traditional LC crossover circuit both deviate, and the resonant points fail to overlap. Neither the amplitude nor the phase of the signals allows HPF and LPF to form an accurate mathematical correspondence. As a result, their amplitudes cannot complement each other, and their phases cannot overlap.\\

\subsection{Resonance peak explanation}
\begin{figure}[h]
  \centering
  \includegraphics[width=0.9\linewidth]{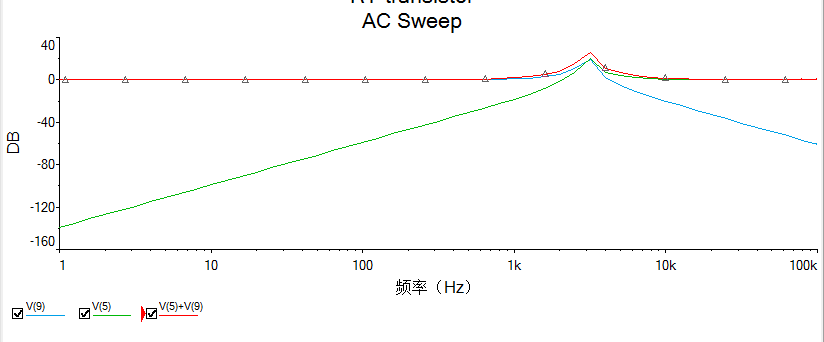}
  \caption{Amplitude of RTR AC-Sweep}
    \label{fig:rtr_ac_sweep}
\end{figure}

As observed in Fig.~\ref{fig:rtr_ac_sweep}, a localized gain peak occurs at the crossover. This is explained by the LC series resonance between primary inductance and effective capacitance. At resonance reactances cancel and net impedance is minimized, producing maximal coupling. Because the phenomenon is reactive, energy is not consumed but temporarily exchanged between inductive and capacitive storage before being delivered to outputs¡ªhence consistent with energy conservation.\\

\subsection{Energy efficiency and robustness}
\begin{figure}[h]
    \centering
    \includegraphics[width=0.7\linewidth]{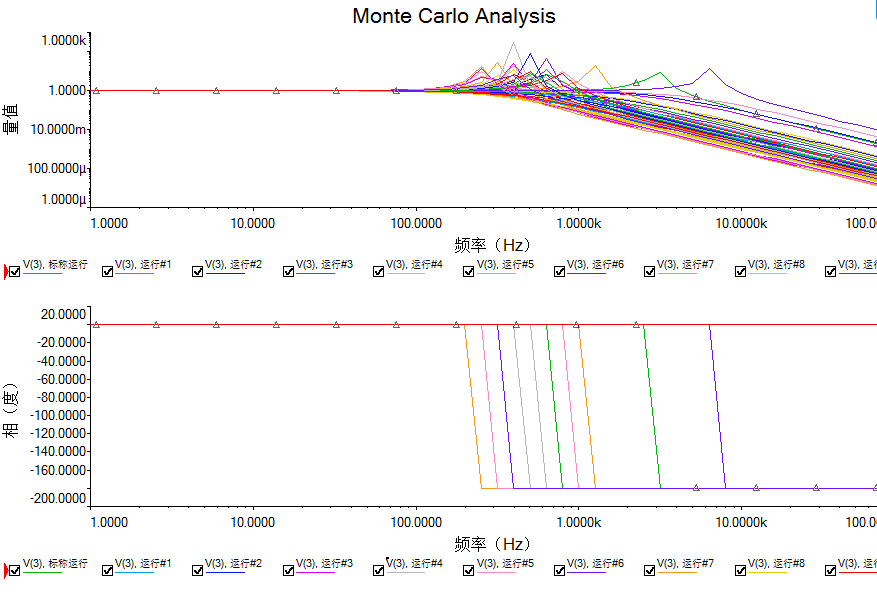}
    \caption{Phase analysis of Monte Carlo / LC crossover}
    \label{fig:monte_carlo}
\end{figure}

Power measurements confirm the sum of LF and HF output powers closely equals input power (insertion loss near 0\,dB). Under Monte Carlo variations with $\pm5\%$ component tolerances, RTR exhibits phase deviations below 1¡ã, As can be seen from Fig.~\ref{fig:monte_carlo}, the phase of the LC filter in the Monte Carlo analysis shows significant instability. These results underscore RTR's robustness to component tolerances.\\

\subsection{Comparison with digital filtering}
Digital FIR/IIR filters are flexible but at high carrier or signal bandwidths require large numbers of taps or high sampling rates, causing heavy computation and latency. RTR performs splitting in the analog domain without DSP overhead, offering real-time, low-latency, and energy-efficient operation particularly advantageous in high-frequency regimes (RF, mmWave).\\

\begin{table}[h]
\centering
\caption{Comparison among LC, FIR/IIR and RTR}
\label{tab:comp}
\begin{tabular}{lccc}
\toprule
Metric & LC & FIR/IIR & RTR \\
\midrule
Insertion loss & 0.3--0.6 dB & none & $\approx$0 dB \\
Phase alignment & poor & can be corrected & near-perfect \\
Latency & none & processing delay & none \\
Tolerance sensitivity & high & low & low \\
Computational cost & none & high & none \\
Energy conservation & partial & partial & yes \\
\bottomrule
\end{tabular}
\end{table}

\subsection{High-frequency splitting advantage}
At very high frequencies, digital filters face two primary challenges: (1) sampling and computation. Nyquist requirements push sampling rates up, increasing required filter taps and processing throughput; (2) latency ¡ª large-tap FIR filters introduce significant group delay that may be unacceptable in real-time systems. RTR bypasses both issues by performing splitting in the analog domain via transformer coupling. The splitting bandwidth is determined by transformer design (windings, core, parasitics) rather than digital clock, allowing RTR to scale into RF and mmWave bands where DSP-based filters become impractical. Consequently, RTR demonstrates superior energy efficiency, lower latency, and stronger robustness in high-frequency splitting tasks.

\section{Applications}
\subsection{HiFi Audio systems}
In high-fidelity loudspeaker and crossovers, RTR ensures phase-coherent outputs for different drivers, improving transient response and stereo imaging. Furthermore, the RTR circuit is particularly well-suited for preamp crossovers because its performance is less susceptible to the complex impedance of loudspeaker loads. By being placed in the pre-amplifier stage, it can more effectively ensure minimal phase deviation after frequency division, leading to a cleaner and more accurate signal reconstruction.

\subsection{Communication front-ends}
RTR can be applied to RF front-ends as a low-loss band splitter, reducing insertion loss and preserving phase coherence required for coherent demodulation.

\subsection{Array signal processing}
For microphone and antenna arrays, RTR preserves inter-channel phase consistency, benefiting beamforming accuracy and sidelobe control.

\section{Network Transmission and Security Applications}
Beyond its traditional applications in analog signal processing, RTR presents compelling opportunities at the physical layer of networked systems:

\subsection{High-speed data demultiplexing}
RTR enables the lossless, real-time demultiplexing of distinct spectral substreams. These substreams may be routed in parallel over separate physical or logical channels, improving aggregate throughput and providing hardware-accelerated demultiplexing.

\subsection{Network security monitoring}
By nondisruptively tapping and isolating a high-frequency data substream (e.g., sideband or HF component), RTR enables real-time deep inspection without interrupting primary data flow. This capability enhances intrusion detection systems (IDS) and traffic analytics with minimal impact on latency and insertion loss.

\subsection{Physical-layer encryption and anti-interference}
RTR can be integrated into schemes where source data is partitioned across complementary LF/HF components (or multiplexed with pseudo-noise carriers). An eavesdropper intercepting a single branch cannot reconstruct the full signal, while authorized receivers reconstruct by combining branches. Such physical-layer splitting can be combined with frequency-hopping or mixing strategies to realize lightweight anti-eavesdropping and anti-jamming defenses.\cite{mukherjee}

\section{Conclusion and Future Work}
We presented RTR, a transformer-based lossless crossover achieving perfect phase alignment and linear complementarity between LF and HF channels. Analytical derivations and circuit simulations confirm RTR's energy-conserving splitting, phase coherence, and robustness to component tolerances. RTR is especially compelling in high-frequency contexts where DSP-based approaches face sampling and latency limits. Future work includes:
\begin{itemize}
  \item Designing tunable RTR topologies (adjustable crossover frequency);
  \item Hardware prototyping and measurement on PCB/bench to validate real-world performance;
  \item Investigating RTR-enabled physical-layer security protocols and multi-path transport architectures, particularly for signal encryption and secure data transmission.
  \item Exploring the application of RTR's fundamental concepts in broader contexts. For instance, similar to how the field of power engineering models frequency division as a continuum to study grid dynamics \cite{Tzounas2022}, our approach could serve as a theoretical basis for exploring new signal processing paradigms in distributed sensor networks and large-scale antenna arrays.
\item Exploring the fundamental concept of RTR in broader signal processing contexts. While our method is based on analog principles, its core idea of precisely managing energy and phase alignment to achieve signal splitting and reconstruction is conceptually similar to advanced digital signal design goals. For example, some digital methods focus on designing signals with specific time-frequency localization properties \cite{Ozdemir2002}. In a similar vein, the RTR's ability to manipulate a signal's frequency and phase in the analog domain could be investigated for applications like signal feature extraction or compression, opening up new avenues for research at the intersection of analog circuit design and digital signal processing theory.
\end{itemize}

\section{Acknowledgments}
The authors express their sincere gratitude to the anonymous reviewers for their insightful comments and constructive suggestions, which were invaluable in significantly improving the quality of this article. 

\bibliographystyle{IEEEtran}

\begin{figure}[h]
  \centering
  \includegraphics[width=0.9\linewidth]{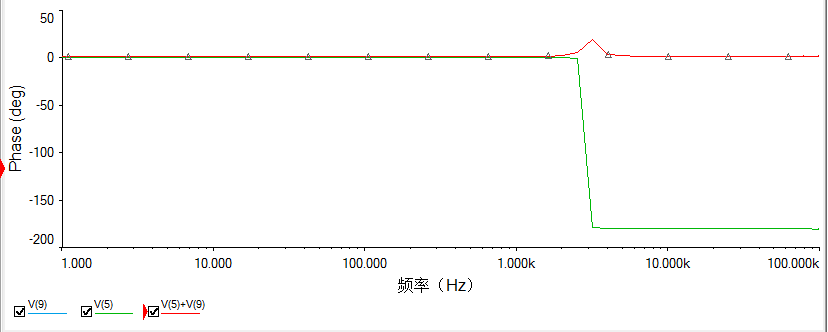}
  \caption{RTR phase responses of LF and HF channels and their sum}
  \label{fig:phase}
\end{figure}

\end{document}